# Database Systems Course: Service Learning Project

Sherri Harms
Department of Computer Science and Information Systems
University of Nebraska at Kearney
Kearney, NE 68849
harmssk@unk.edu

## Abstract

This paper describes a service learning project used in an upper-level/graduate-level database systems course. Students complete a small database project for a real client. The final product must match the client specification and needs, and include the database design and the final working database system with embedded user documentation. The solution must be implemented in a way to make it as easy to use as possible for the client. Students are expected to conduct professional meetings with their clients to understand the project, analyze the project's requirements, as well as design and implement the solution to the project. Students must have each milestone approved before starting the next phase of the project.

The student learning objectives of a database system semester project are to: analyze a client's information system problem and determine the requirements for the solution; design a suitable database solution to the problem; use software design and development tools to design and develop a solution to the problem; communicate and interact with a client on a professional level; prepare effective documentation for both non-technical and technical software users; and interact ethically with all persons involved with a project. The broader impact objectives of a database system semester project are to: provide needed database solutions for organizations and businesses in the local area; provide a resume and portfolio-building opportunity for the students; provide a measure for assessing how well the program meets it mission; provide a mechanism for implementing service-based learning; provide a mechanism for outreach to local-area organizations and businesses; and provide a starting-point for undergraduate research projects.

# 1 Introduction

This paper describes a service learning project used in an upper-level/graduate-level database systems course (CSIS 425/825P Database Systems) offered by the Computer Science and Information Systems Department at the University of Nebraska at Kearney (UNK). This course is offered as a blended on-campus/on-line course. Class lectures are recorded and the Blackboard learning content management system is used to distribute course materials and to accept course assignments. The overall student learning objectives of this course are to:

1. Design and implement a simple database system from user's requirements.
2. Understand various data models used by database management systems.
3. Apply rules of normalization to develop a fully normalized data model.
4. Use relational database management systems to implement data models.
5. Understand multiple-user database processing issues.
6. Analyze and implement system requirement changes.
7. Understand enterprise database processing issues and implement a basic web-based database system.

# 2 Project Description

To achieve these student learning objectives, a semester-long independent project is used. This project is a small database project for a real client. The final product must match the client specification and needs, and include the database design, and the final working database system with embedded user documentation. The solution must be implemented in a way to make it as easy to use as possible for the client. Students are expected to conduct professional meetings with their clients to understand the project, analyze the project's requirements, as well as design and implement the solution to the project. Students must have each milestone approved before starting the next phase of the project.

The student is expected to come up with their own semester project, but we have a list of projects that are needed by offices on campus or organizations from the local-area. These projects provide a service to the community as well as meet the student learning objectives [1-3].

Sample projects from fall 2011 include: a database system to manage driver's education student records; a dataset system to record a local-area children's Awana Club attendance and books read per student; database systems to manage information regarding student's completion of Advanced Driving Technique (ADT) classes, Motorcycle Safety Education (MSE) classes and Safety Training Option Program (STOP) training; a database system to track laptop usage and damage for a local-area school system; a database system to track a community theatre's operation; a database system to record a wind turbine's data regarding wind speed and energy produced on an hourly-basis; an online client management system for a local-area law office; a database system to manage construction jobs and costs for a local-area construction company; and a database system to manage a trucking company's expenses and profits per trip/load.

## 2.1 Project Objectives

The student learning objectives of a database system semester project are to:
1. Analyze a client's information system problem and determine the requirements for the solution.
2. Design a suitable database solution to the problem.
3. Use software design and development tools to design and develop a solution to the problem.
4. Communicate and interact with a client on a professional level.
5. Prepare effective documentation for both non-technical and technical software users.
6. Interact ethically with all persons involved with a project.

The broader impact objectives of a database system semester project are to:
1. Provide needed database solutions for organizations and businesses in the local area.
2. Provide a resume and portfolio-building opportunity for the students.
3. Provide a measure for assessing how well the program meets it mission.
4. Provide a mechanism for implementing service-based learning.
5. Provide a mechanism for outreach to local-area organizations and businesses.
6. Provide a starting-point for undergraduate research projects.

## 2.2 Project Milestones

The milestones in this project include the project proposal; a meeting with the instructor to discuss the project design; the project design; the project tables and relationships; the project working prototype; and the final completed project with the client's evaluation. Appendix A shows the final completed project grading rubric. Appendix B shows the client evaluation form. Appendix C shows documentation provided to the students.

### 2.2.1 Project Proposal

By the end of the third week of class, students must submit their project proposals. The proposal should briefly describe the project the student plans to work on for the semester and include the client's contact information. The instructor may require changes to the proposal. Thus, the proposal may need to be submitted several times until it meets the satisfaction of the client, the student and the instructor. One item that the instructor must manage is the scope of the project. Because the students are not experts in managing scope, it is the instructor's responsibility to help the students work with the client to develop a manageable project and to not let the scope grow over the semester.

### 2.2.2 Project Design

During weeks 6-8, each student is required to schedule a half-hour appointment with the instructor (via virtual meeting or face to face), to discuss the student's database design and the project's requirements. The student must show and describe the project's design, which means the student must have met with his/her client and must understand the project requirements.

Each student is also required to have two other students review their design and to review two other students' designs. Students are required to document what they recommended for the designs they reviewed as well as document what the other students discussed/recommended to them. This gives students more experience with designing projects and also helps them develop their technical communication skills.

By the end of the eighth week of class, the students must submit their design plans. The business rules must be documented and included as part of the design plans. The student review information, with classmate names must also be included. The students submit a Microsoft Visio [4] file with their ERD design and a document with the business rules & classmate suggestions. The instructor or client may require changes to the design. Thus, the design may need to be submitted several times until it meets the satisfaction of the client, the student and the instructor.

### 2.2.3 Project Tables & Relationships

Once the design has been approved by the instructor and the client, the student starts to work on the implementation. My guiding principle is for the project implementation to have five to eight tables. Projects that require more than that are difficult to manage for the student. Students are asked to work with the client to scale back projects, if needed, so that no more than eight tables are implemented.

By the end of week ten, each student must submit their database, with the tables and relationships defined. They must include at least 5 sample records in each table. The students must also submit a final ERD as well. The instructor may require changes to the table and relationship implementation – particularly with foreign-key implementations. Thus, the tables and relationship implementation may need to be submitted several times until it meets the satisfaction of the client, the student and the instructor.

### 2.2.4 Project Prototype

By the end of week 13, each student must submit their working database prototype solution, with forms, reports, menus, embedded documentation, etc. The students are required to provide their prototype solution to their clients as well.

The instructor and client provide detailed feedback regarding necessary and suggested changes to the database system. Additionally, the client must complete a client feedback form (included in Appendix B), and return it directly to the instructor.

### 2.2.5 Final Project with Documentation

By the end of week 15, each student must submit their final complete working database solution with embedded documentation. The final project grading rubric is included in Appendix A. The database system should have the look and feel that matches the client's needs. It should have an opening page that provides a portal to all functionality in the system. The daily use activities should be easy to find and complete. The necessary use activities should be easy to find and have guides to completion. The ad hoc uses should be easy to find and have guides to completion. Create, Read, Update, Delete (CRUD) requirements must be met. (All data collected is being used; all needed data has a way to be easily stored and updated.)

The embedded documentation should guide the users through the uses of the database system. Students are to use screen images with text explanations on how to add records, modify records, find records, etc. Student must include their final ERD and any notes to explain unique design and implementation items in their database solution. This includes items such as which query was used for a given form or report, special input requirements into a field, using combo boxes that allow for new items to be added to the list, etc.

# 4 Sample Project

To illustrate the kinds of projects students complete in this course, a sample project is described below. This project is for a small community theatre. The database keeps track of theatre members' contact info; patron levels and ticket numbers; volunteer participation; mailing lists and patron lists; attendance of patrons; past participation of patrons; and special donations. It has the ability to prepare reports of the above data for the theatre board (mostly yearly totals; print labels; and to receive/send info via emails.

## 4.1 Sample Project Design

### 4.1.1 Sample Business Rules

The business rules that the student discovered following the initial interview with the client started with the rule that members are generally considered as customers or people who give donations (which are also generally customers). More specifically members are anyone who is in the mailing list. Any member can give a 'special donation' which is not considered as a patron donation. Also, members can volunteer. A patron is a member who donates a certain amount to the theatre and receives tickets based on his/her donation level. Patrons' past participation needs to be recorded along with the patrons' current participation.

### 4.1.2 Sample Follow-up questions for the client

After establishing the business rules for this project, the student had the following questions that he felt needed to be addressed before the design could be completed:

1) <u>Special Donation</u> – For description, does the client want to look up a certain type of donation?
2) <u>Volunteer Activity</u> – For description, does the client want to look up a certain activity?
3) <u>Patron</u> – do only the dates of attendance need to be stored, or should there be records for shows?
4) <u>Patron</u> – how should a patron's previous participation be stored?

### 4.1.3 Sample Project Entity Relationship Diagram (ERD)

After resolving these issues, and after meeting with both the client and the instructor, the entity relationship diagram (ERD) shown in Figure 1 below was developed. While nine entities on an ERD are more than my suggested number of entities to implement, Patron Level is a look-up table which has data that will not change over time.

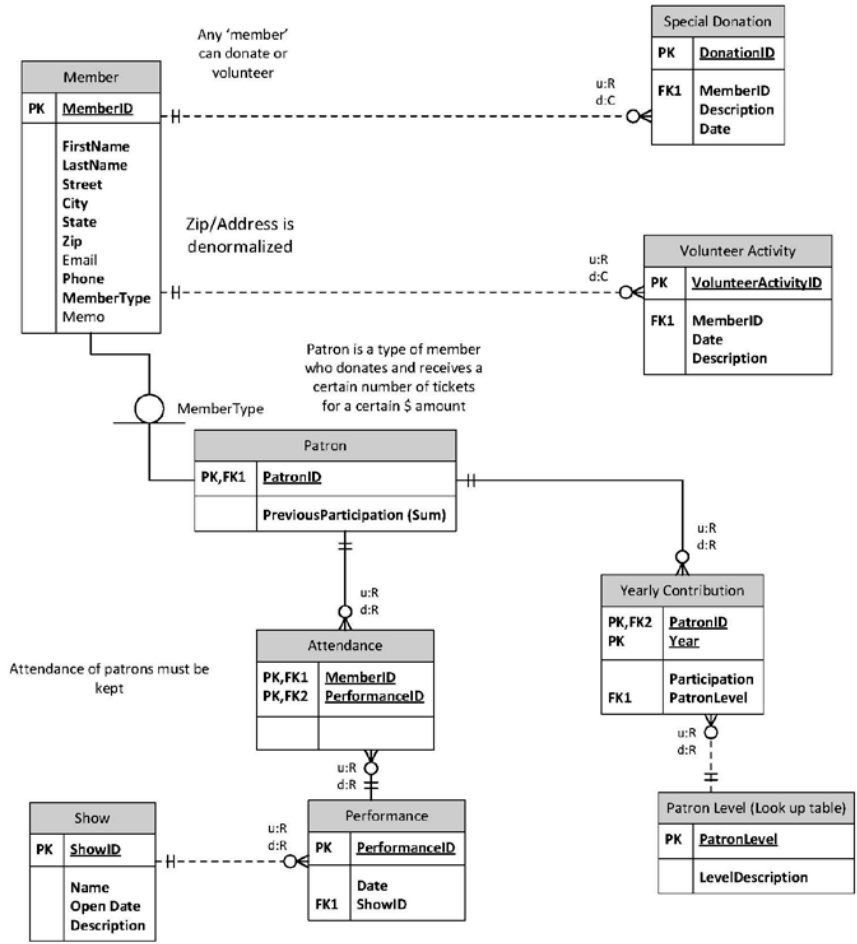

Figure 1. Sample Project ERD

### 4.1.4 Sample Student Reviews of the project design

When this student reflected on the student reviews of his project, he indicated that "I think everyone who reviewed my project questioned the patron/attendance entity relationship. I agree with what they had questions about. I still need to change a few things in my Visio diagram after I discuss more details with my client." Below are the student reviews for the sample project.

Student 1 review:
*I was a little confused with the Attendance and Showing relationship. Are you counting a showing as a unique show? Can the same show be shown multiple times, thus having more than one showing? If the same show can be shown multiple times, I would suggest*

*adding another table named 'Show' that relates to the 'Showing' in a 1 - many relationship. The Show table would have the show's name, description, and an Id primary key. The 'Showing' table would be a go between, just having an Id primary key, Date, and the Id of the Show. Of course, that depends on whether if you are counting each showing as a unique show.*

Student 2 review:
*For the Patron attendance, should there be a date of the meeting or if they show up? Are they meant to contact you so they specifically get mentioned? Otherwise, your entities are very good!*

### 4.1.5 Sample Project Tables and Relationships

From the ERD shown in Figure 1, the tables with sample data shown in Figure 2 were implemented in Microsoft Access [5]. By inserting sample data into the tables, the student has a better understanding of the project prior to building the interface.

**Member:**

| MemberID | FirstName | LastName | Street | City | State | Zipcode | Email | Phone | Patron | Memo |
|---|---|---|---|---|---|---|---|---|---|---|
| 1 | Test | Testing | T Street | T City | TS | 65382 | test@test.com | 2038480248 | TRUE | |
| 2 | Second | Person2 | 45w4k | New City | ST | 98583 | | | TRUE | |
| 3 | Third | Person3 | kdjf | djkfak | NW | 84394 | | | TRUE | |
| 4 | adsjk | kdjfk | adjf | kjdf | JK | 83948 | | | TRUE | |
| 5 | bkgjs | dkjf | kdjf | ksjdf | HD | 39409 | | | TRUE | |
| 6 | sksjd | jksdjf | kjdf | kjkj | JS | 34820 | | | FALSE | |

**Special Donation:**

| DonationID | Member | DonationDate | DonationDescription |
|---|---|---|---|
| 1 | 1 | 11/1/2011 | Donated something |
| 2 | 2 | 10/18/2011 | Donated different item |
| 3 | 4 | 11/16/2011 | Blah Blah |
| 4 | 5 | 11/11/2011 | New |
| 5 | 2 | 11/19/2011 | Another Donation |

**Volunteer Activity:**

| VolunteerActivityID | Member | ActivityDate | ActivityDescription |
|---|---|---|---|
| 1 | 1 | 10/26/2011 | Did something that can't be ca |
| 2 | 3 | 10/13/2011 | Something else |
| 3 | 4 | 11/17/2011 | New |
| 4 | 4 | 11/25/2011 | new1 |
| 5 | 2 | 11/4/2011 | new2 |

**Patron:**

| Member | PreviousParticipation |
|---|---|
| 1 | 384 |
| 2 | 52 |
| 3 | 582 |
| 4 | 482 |
| 5 | 2,809.00 |

**Yearly Contribution:**

| Patron | Contribution | Participation | PatronLevel |
|---|---|---|---|
| 1 | 10/11/2011 | 23.00 | 1 |
| 2 | 11/19/2011 | 180.00 | 3 |
| 3 | 11/18/2011 | 84.00 | 2 |
| 4 | 11/18/2011 | 32.00 | 1 |
| 5 | 11/4/2011 | 10.00 | 1 |

**Attendance:**

| Member | Performance |
|---|---|
| 1 | 1 |
| 1 | 2 |
| 2 | 1 |
| 3 | 2 |
| 4 | 5 |

**Performance:**

| PerformanceID | Show | PerformanceDate |
|---|---|---|
| 1 | 1 | 11/30/2011 |
| 2 | 1 | 11/17/2011 |
| 3 | 2 | 11/27/2011 |
| 4 | 3 | 11/18/2011 |
| 5 | 5 | 11/14/2011 |

**Show:**

| ShowID | ShowName | OpenDate | Description |
|---|---|---|---|
| 1 | 1st Show | 11/2/2011 | First |
| 2 | 2nd Show | 11/17/2011 | Second |
| 3 | 3rd Show | 11/18/2011 | |
| 4 | 4th Show | 11/11/2011 | |
| 5 | 5th Show | 11/12/2011 | |

**Patron Level:**

| PatronLevel | LevelDescription |
|---|---|
| 1 | certain amount… |
| 2 | another amount |
| 3 | last amount |

**Figure 2. Sample project tables**

### 4.1.6 Sample Project Screen Shots

Sample screen shots from the final sample project, implemented in Microsoft Access are shown in Figures 3-5. The opening menu shown in Figure 3 provides buttons to enter/modify data, print labels and print reports. If the user chooses to enter/modify data, the user is shown the options for entering/modify the tables independently, as shown on Figure 4. Figure 5 shows the form for entering/modifying a member's information. Notice the "?" on the menu pages, which provides the documentation for the system. Figure 6 shows one of these documentation pages. For consistency, all data entry forms had the same menu structure.

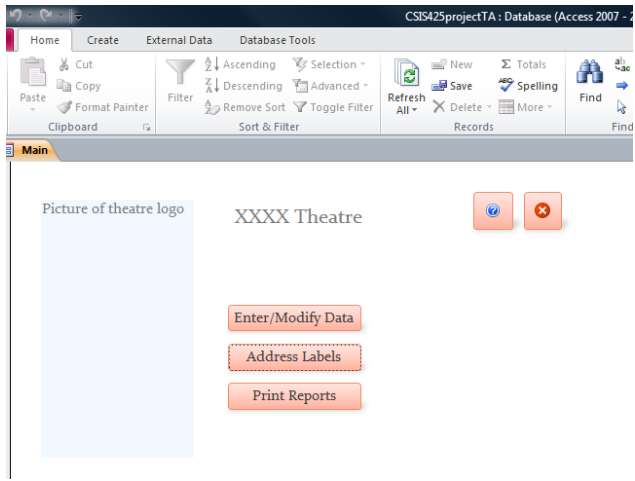

Figure 3. Theatre Opening Menu

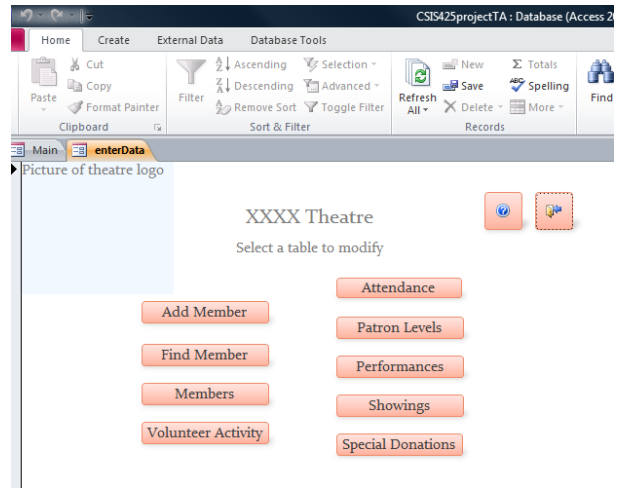

Figure 4. Enter/Modify Data Menu

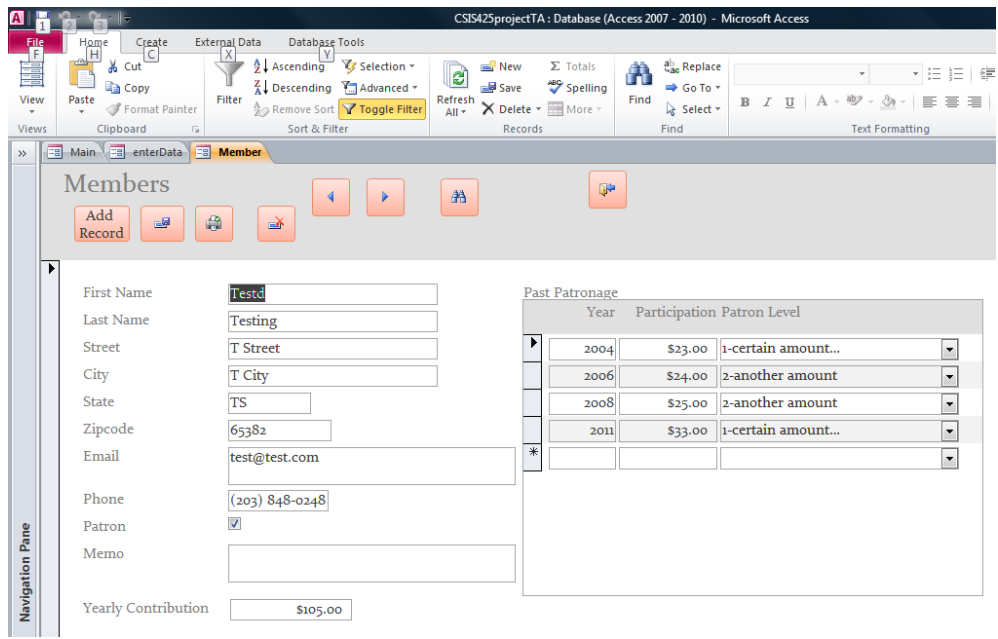

Figure 5. Member Entry Form

# 5 Discussion of the Success of this Project

I have been using this project in my database class for the past nine years. The quality and quantity of projects that student have developed for UNK campus offices and student organizations, such as the International Student Program, Residential Life Program,

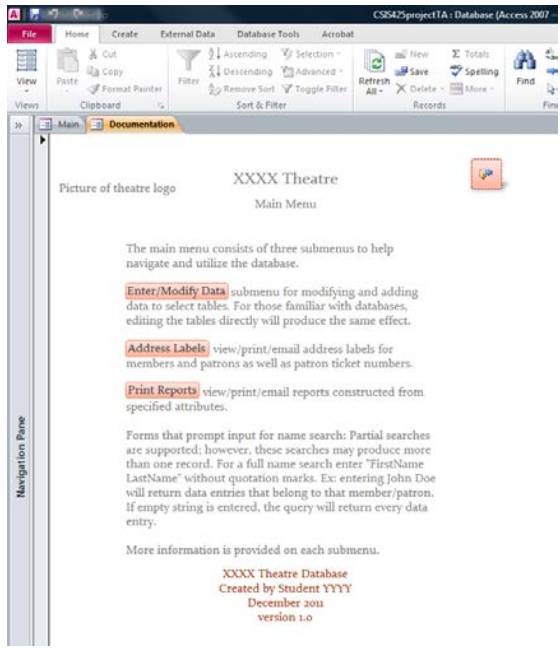

**Figure 6. Sample Documentation Page**

Student Support Services, UNK Finance Office, and UNK Safety Center has been excellent. I am also pleased of the projects they have created for organizations such as Mid-Nebraska Community Action Center, Minority Health Clinic, Christian Heritage group homes for children, Kearney Community Theatre, Chrysalis Christian Retreats, and Habitat for Humanity; and for local area small businesses. Local organizations, offices, and businesses contact me throughout the year to "get in line" to have a database system developed for their needs.

This project forces students to apply what they learn in the classroom to a real world problem that is often messy. It provides students with opportunities to develop their professional communication skills, which are critical to today's technology graduates. It requires students to interact with clients in an ethical manner as well, as clients often must disclose sensitive information for the student to complete the project.

Using real-world projects requires the instructor to manage the course differently than a normal lecture course. Making sure that students meet each milestone before continuing on to the next step requires re-grading and reviewing each student's work numerous times. While this is time consuming, students learn from the iterative process, and they are motivated to complete each milestone on-time. Earlier milestones are weighted less than the final implementation, to allow for students to modify and re-work the design and prototype. In total, this project is worth 35% of the grade for the course.

I have learned that it is important to have the one-on-one meeting with each student to discuss their design early in the semester I have found that seeing the tables and relationships implemented correctly before the student starts to implement the user interface (forms, reports, web pages, etc.) is also important. I was surprised how many students had difficulty converting their design into working tables, with appropriate keys, foreign keys, field sizes, etc. It is also important to have the working prototype completed during the 13$^{th}$ week to allow time for students to implement the corrections suggested by the client and myself. Without this time for corrections, the projects often were "almost usable". Unfortunately, an information system that is "almost usable" is in fact, not usable at all. This would often leave the client/student relationship on bad terms. By moving the deadline up by two weeks and providing detailed feedback to the student for necessary corrections, the projects more often than not, are now readily used projects.

Fall 2011 was the first semester this project was used in a blended deliver course. I was pleasantly surprised at how well this project worked for on-line students. Finding time to meet with on-line students required flexibility on my part (such as an evening meetings via virtual meetings), but on-line students were successful at working with their clients and meeting milestones. Six lecture hours are dedicated to project milestone lab days throughout the semester. Since on-line students are not able to go to the lab, I created several tutorials for completing the milestones, using the Camtasia Studio [6] software and I provided students with Atomic Learning [7] on-line tutorials. All students found these to be useful. I also provided the students with Microsoft MSDN Academic Alliance [7] accounts, for educational use of Microsoft Visio and Access.

Occasionally students choose to complete projects for a family member's or friends' business. One father commented on the client evaluation of his son's project that he was "finally able to see some results from his son's education." While there is concern that the family member or friend will be overly generous in the evaluation of the project, I have found them to be quite honest in the project evaluation. Additionally, I complete my own evaluation of the project. Another advantage of completing a project for a family member or a friend is that the student will likely stay in contact with the client and be available to make changes over time.

Alumni often respond on surveys that the database project was one of the most valuable projects during their time as a student. They indicate that they were able to use that project on their resume and in interviews as an example of work they completed.

We also use these projects as part of our departmental assessment. In fact, the client evaluation is one of the assessment measures used to evaluation how well the department is meeting its mission. The department reports assessment results every three years.

Finally, there is at least one student project per semester that is continued in some manner as an undergraduate research project. These projects have been presented at our university student research day and at national conferences, such as the National Conference on Undergraduate Research (NCUR).

## 6 Conclusion

This paper described a semester long hands-on project for an upper-level/graduate-level database class. Students learn to analyze a client's information system problem and determine the requirements for the solution; design a suitable database solution; use software design and development tools; communicate and interact with a client on a professional level; prepare effective documentation; and interact ethically with all persons involved with a project. This project provides: needed database solutions for organizations and businesses in the local area; a resume and portfolio-building opportunity for the students; a measure for assessing how well the program meets it mission; a mechanism for implementing service-based learning; a mechanism for outreach to local-area organizations and businesses; and a starting-point for undergraduate research projects.

# Appendix A. Final project implementation grading rubric

Student _______________________________    Project Name: ___________________

| Criterion | Possible Points | Points Earned |
|---|---|---|
| **1. Functionality** | | |
| User interface<br>    Organization (Menus, forms, reports, buttons, layout, background)<br>    Look and feel (Match client's design) | 25 | |
| Degree to which it works as specified<br>    Daily use activities – easy to find and complete<br>    Necessary Use – easy to find and guides to completion<br>    Ad hoc uses - easy to find and guides to completion<br>    Activities work properly & give correct results | 50 | |
| **2. Design** | | |
| Design – ERD, I/O design, forms, reports<br>Consistent, matches client's requirements | 25 | |
| CRUD requirements met<br>(All data collected is being used; All needed data is being stored and updated) | 10 | |
| Ease of Use/Self Explanatory/Documentation | 10 | |
| **3. Client Evaluation** | **30** | |
| Total Possible | 150 | |

**Comments:**

# Appendix B. Final project client evaluation form
# COMPUTER SCIENCE AND INFORMATION SYSTEMS
# DATA BASE SYSTEMS CLIENT SURVEY

Thank you for agreeing to work with a team of students in the Database Systems Class at UNK. As a part of our assessment plan we would like your opinions concerning your experience with our students. Your opinions are valuable to us. Please complete this questionnaire and email it to harmssk@unk.edu, or mail it to: Sherri Harms, Department of Computer Science and Information Systems, University of Nebraska-Kearney, Kearney, NE 68849
Thank you!!

**Student Name** \_\_\_\_\_\_\_\_\_\_\_\_\_\_\_\_\_\_\_\_\_\_\_\_\_\_\_\_\_\_\_\_\_\_\_\_\_\_\_\_\_\_\_\_\_\_\_\_\_\_\_\_\_\_\_\_\_\_

**Project Name/Description** \_\_\_\_\_\_\_\_\_\_\_\_\_\_\_\_\_\_\_\_\_\_\_\_\_\_\_\_\_\_\_\_\_\_\_\_\_\_\_\_\_\_\_\_\_\_

**Evaluator** \_\_\_\_\_\_\_\_\_\_\_\_\_\_\_\_\_\_\_\_\_\_\_\_\_\_\_\_\_\_\_\_\_\_\_\_\_\_\_\_\_\_\_\_\_\_\_\_\_\_\_\_\_\_\_\_\_\_\_\_

Please circle one response for each statement indicating the degree with which you agree with the statement.

The students I worked with in the Database Systems project:

| | | | | | |
|---|---|---|---|---|---|
| Effectively analyzed the problem and determined the requirements for a solution. | Agree Strongly | Agree | Neither agree nor disagree | Disagree | Disagree Strongly |
| Designed a suitable database solution to the problem. | Agree Strongly | Agree | Neither agree nor disagree | Disagree | Disagree Strongly |
| Used software design and development tools to design and develop a solution to the problem. | Agree Strongly | Agree | Neither agree nor disagree | Disagree | Disagree Strongly |
| Communicated and interacted with us on a professional level. | Agree Strongly | Agree | Neither agree nor disagree | Disagree | Disagree Strongly |
| Prepared effective documentation for both non-technical and technical software users. | Agree Strongly | Agree | Neither agree nor disagree | Disagree | Disagree Strongly |
| Given a suitable job opening, I would be willing to employ the students with which I worked. | Agree Strongly | Agree | Neither agree nor disagree | Disagree | Disagree Strongly |
| Interacted ethically with all persons involved with the project. | Agree Strongly | Agree | Neither agree nor disagree | Disagree | Disagree Strongly |
| The grade I would assign to these students is: | A | B | C | D | F |

Comments:

# Appendix C. Guidelines for Completing a Database System Project

## C.1. Initial Investigation Stage - Discussion with your Client

- First Meeting Goals:
  - Have the client explain their overall purpose and goals of the database, and how the database will be used (in general).
  - Listen for nouns in their speech (used for possible entities).
- Second Meeting Goals:
  - Listen for attributes (such as possible lists) – but usually they do not need to elaborate on these in this interview.  (This may be part of the first meeting.)
  - Listen for verbs that define relationships - "Is a"(subtype), "has a", "by", and other verbs that "tie" the nouns together. For example:  We keep track of the student by major.  Each faculty teaches many classes.
- Other stage one goals:
  - Get a general idea of their interface needs (input & output needs) – Is it web-based?  Is it a single person system?  Is security an issue?  Are standard reports needed?  Are ad-hoc queries needed?
  - Find out what DBMS will work best for their needs (Access, MySQL, Oracle, other?)  They probably will not know what DBMS will work best for their needs, so based on their discussion you will need to make a recommendation.

## C.2. Design Stage

- Identify entities. Plural words can be either a list (a separate entity), or a description attribute.
- Identify attributes (they describe the entity's characteristics)
- Identify identifiers
- Identify/create relationships
- Follow the normalization process to make sure there are no repeating values in any field and make sure all attributes are dependent on the whole primary key.

- Meetings in this stage:
  - Bring your rough ERD to help the client visualize the design with you.
  - Questions about the entities:  Ask the client to tell you more about each of the entities you have discovered.   What data do they keep for each entity?  How much detail do they need? How will this data be used? How are the entities "tied" together (if you have not discovered that earlier)?
  - Questions about the attributes:  Ask to see sample data, ask about the valid range of values, and other constraints on the domains of each attribute.
  - Questions about the relationships: Ask the client to tell you more about each relationship.  How many instances of each entity type are in the relationship, is it an optional relationship?

- Have two other students review your design.
- Meet with the instructor and your client to make sure your design is correct.

### C.3. User Interface Design

- Develop your interface prototype and review with your client. Ask to see your client's existing forms and reports, and/or have them describe their "ideal" interface design (even if these are hand drawn.)

- Menus
    o Your interface should have an opening window/menu which includes access to all forms/reports. This menu should open when the project is opened.
    o Make the "daily use" activities easy to get to.
    o Provide the "necessary use" activities with extra help/guidance – especially if these happen infrequently.
    o Provide ways to perform ad hoc queries.

- Forms
    o When creating a form over a table, make sure the layout and "tab" ordering follows the natural progression of the form.
    o Use appropriate labels, which are likely to be different than the field name.
    o Use subforms to enter related child data.
    o Use consistent buttons on all data entry forms.

- Reports
    o Used to view data in meaningful ways.
    o Usually built over a query, often times this is a parameterized query.
    o Set report properties appropriately, and be consistent between all reports.

- Functionality is important
    o It must meet the client needs specified in the proposal
    o Your project needs to have ways for the user to:
        - Create – add new data to all tables & all fields
        - Read – view all data in meaningful ways
        - Update – modify all existing data
        - Delete – delete existing data – with care!
    o By using different buttons on the main menu, with slightly different functionality, you can implement the needed CRUD activities for your project.

- Organization, design, and ease of use are important
    o Submenus are useful
    o Consistent placement of buttons is helpful
    o The look and feel should reflect your client's needs. (This includes background, images, fonts, etc.)
    o The user should quickly and easily become familiar with the system.
    o Provide an embedded user manual or help buttons to explain.

- Helpful extra features (for MS Access projects)
    o Allow the user to automatically add a new value to a lookup table without leaving the data entry form.

o Use the calendar feature for entering or updating a date field.
   o Automatically set a remaining balance when a transaction is started using the initial amount due.
   o Automatically update a remaining balance after a payment is made.

**C.4. Implementation**

- Build from your prototype to develop the interface. Test and retest the functionality. Have the client, the instructor, and another student test the prototype.

**Each step:**
- Keep your client informed. When in doubt, ask them – it is their system.
- During your meetings with them, the client should do 80% of the talking – your job is to listen and take brief notes or sketches. After the meeting, expand your notes. Always thank your client for their assistance and support of your project. Keep your meetings on track – time is valuable to you and your client.
- Help your client visualize the final product at each step. The focus will become clearer each step, as it also becomes more detailed.
- Remember your clients know their business and have a good understanding of their data needs, but they should not be expected to know data modeling (and you are not expected to teach them). Use their lingo, that is, words that have meaning to them.
- Remind the user's to think "outside the box" & plan for the future – will this design meet their future needs? Are there other ways that they plan to use the data, i.e., do they sometimes need to export the data to Excel for statistical processing? If so, are there ways to incorporate this into the system

# References

[1] Bringle, R.D., Hatcher, J.A., Implementing Service Learning in Higher Education, *The Journal of Higher Education*, 67(2): 221-239, (Mar. - Apr., 1996), Article Stable URL: http://www.jstor.org/stable/2943981.

[2] Sanderson, P. Where's (the) computer science in service-learning? *J. Comput. Small Coll.* 19 (1): 83-89, October 2003.

[3] Tan, J., Phillips, J. Incorporating service learning into computer science courses. *J. Comput. Small Coll.* 20 (4): 57-62, April 2005.

[4] Microsoft Visio. http://office.microsoft.com/en-us/visio, February 2012.

[5] Microsoft Access. http://office.microsoft.com/en-us/access, February 2012.

[6] Camtasia Studio software, http://www.techsmith.com/camtasia.html, February 2012.

[7] Atomic Learning., http://www.atomiclearning.com, February 2012.

[8] Microsoft MSDN Academic Alliance. https://www.dreamspark.com, February 2012.